\documentclass[aps,prb,final,twocolumn,nobibnotes,showpacs,showkeys,floatfix]{revtex4}

\usepackage{amsmath}
\usepackage{graphicx}
\usepackage{amssymb,amsfonts}
\usepackage{bm}
\usepackage{dcolumn}
\usepackage[hypertex]{hyperref}

\newcommand{\unit}[1]{\ensuremath{{\hat{\boldsymbol{#1}}}}}
\newcommand{\der}[3][]{\frac{d^{#1} #2}{d {#3}^{#1}}}
\newcommand{\pder}[3][]{\frac{\partial^{#1} #2}{\partial #3^{#1}}}
\newcommand{\del}{\boldsymbol{\nabla}}
\newcommand{\grad}{\del}
\newcommand{\cross}{\times}

\newcommand{\curl}[1]{\del\cross{ \bf #1 }}

\begin{document}

\title{Measurements of the magnetic field induced by a turbulent flow
of liquid metal}

\author{M.~D. Nornberg}
\author{E.~J. Spence}
\author{R.~D. Kendrick}
\author{C.~M. Jacobson}
\author{C.~B. Forest}
\email{cbforest@wisc.edu}
\thanks{Invited speaker}

\affiliation{Department of Physics\\ University of Wisconsin-Madison\\
1150 University Ave.\\ Madison, WI 53706}

\date{December 23, 2005}

\begin{abstract}
Initial results from the Madison Dynamo Experiment provide details of
the inductive response of a turbulent flow of liquid sodium to an
applied magnetic field. The magnetic field structure is reconstructed
from both internal and external measurements. A mean toroidal magnetic
field is induced by the flow when an axial field is applied, thereby
demonstrating the omega effect.  Poloidal magnetic flux is expelled
from the fluid by the poloidal flow.  Small-scale magnetic field
structures are generated by turbulence in the flow. The resulting
magnetic power spectrum exhibits a power-law scaling consistent with
the equipartition of the magnetic field with a turbulent velocity
field. The magnetic power spectrum has an apparent knee at the
resistive dissipation scale. Large-scale eddies in the flow cause
significant changes to the instantaneous flow profile resulting in
intermittent bursts of non-axisymmetric magnetic fields, demonstrating
that the transition to a dynamo is not smooth for a turbulent flow.
\end{abstract}

\keywords{magnetohydrodynamics, MHD, dynamo, turbulence}
\pacs{47.65.+a, 91.25.Cw}
\preprint{POP29524DPP05A}

\maketitle

\section{Background}
\label{sec:background}

The generation of magnetic fields by flowing electrically-conducting
fluids is a long-standing problem in plasma physics, astrophysics, and
geophysics. Of particular interest is the role of turbulence in either
supporting, or suppressing, magnetic fields.  Dynamos have been
created in the laboratory by driving helical flows of liquid metal
through pipes.\cite{stieglitz:561,gailitis:4365} Measurements of the
onset of magnetic field growth agree with predictions from laminar
theory, suggesting that turbulence played little role in these
experiments. The pipe geometry, however, limits the development of
eddies to the width of the pipe, a scale much smaller than the device
size, thereby inhibiting dynamics due to large-scale turbulence. Since
astrophysical flows lack this scale separation, understanding the role
of large-scale turbulence is especially important in modeling the
dynamo. The Madison Dynamo Experiment was built to characterize these
dynamics.

\begin{figure*}
\includegraphics{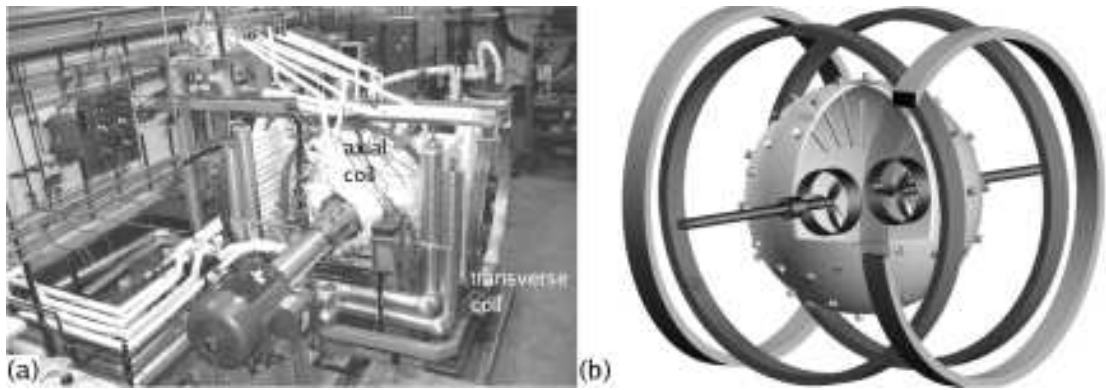}
\caption{(a) Photograph and (b) schematic of the Madison Dynamo
Experiment. The sphere is 1\,meter in diameter. It is filled with
$105$--$110^\circ$C liquid sodium, and a flow is created by two
counter-rotating impellers. Two sets of coils, one coaxial with and
one transverse to the drive shafts, are used to apply various magnetic
field configurations. The magnetic field induced by the flow is
measured using Hall-effect sensors both on the surface of the sphere
and within tubes that extend into the flow.}
\label{fig:experiment_schematic}
\end{figure*}

The experiment, shown in Fig.\,\ref{fig:experiment_schematic}, uses
two impellers to generate a double-vortex flow of liquid sodium in a
1\,m diameter spherical vessel.  Liquid sodium is used because of its
high conductivity ($\sigma = 10^7\,\Omega^{-1}\mbox{m}^{-1}$). The
choice of a spherical geometry is motivated by the computational work
of Dudley and James, which demonstrated that simple time-stationary
vortices can generate magnetic fields at relatively low
speeds.\cite{Dudley_and_James_PRSLA_1989} The flow has been measured
in hydrodynamic experiments and is predicted by laminar dynamo theory
to generate a transient magnetic field\cite{OConnell_et_al_2000} by a
slow-dynamo mechanism that stretches and twists magnetic field lines
to regenerate the initial seed field.\cite{Forest_et_al_MHD_2002} Due
to the low kinematic viscosity ($\nu = 7 \times 10^{-5}\ {\rm
m}^2/{\rm s}$), and hence low Prandtl number of liquid sodium
(\mbox{$Pr \equiv \mu_0\sigma\nu = 9 \times 10^{-4}$}), and the large
magnetic Reynolds numbers required for spontaneous field growth
(\mbox{$Rm \equiv \mu_0 \sigma L V_0 \sim 10^2$}), the flows required
for a dynamo are extremely turbulent (\mbox{$Re \equiv L V_0/\nu =
Rm/Pr \sim 10^5$}). These types of flows are also studied in
experiments at Maryland\cite{Lathrop_and_Shew_and_Sisan_PPCF_2001} and
Cadarache.\cite{bourgoin:3046} The open geometry of these experiments
provides the opportunity to study the role of large-scale turbulence
in magnetic field generation.

The purpose of this correspondence is to provide a description of the
Madison Dynamo Experiment and to report the initial measurements of the
magnetic field induced by the flowing liquid metal.  The theory of
laminar kinematic dynamos and hydrodynamic measurements of the flow
are reviewed in Sec.\,\ref{sec:kinematicdynamo}.  The experimental
apparatus and its diagnostics are described in
Sec.\,\ref{sec:experiment}.  Measurements of the mean magnetic field
induced when an external magnetic field is applied to the turbulent
flow are compared to predictions from the laminar theory in
Sec.\,\ref{sec:magneticfield}. Measurements of the magnetic field
fluctuations are used to study the properties of MHD turbulence in
Sec.\,\ref{sec:turbulence}.

\section{The Kinematic dynamo}
\label{sec:kinematicdynamo}

\subsection{Theory}

The evolution of the magnetic and velocity fields in an incompressible
conducting fluid is governed by the magnetic induction equation and
Navier-Stokes equation with a Lorentz forcing term:
\begin{eqnarray}
\pder{ \mathbf{B} }{ t } & = & \curl{ \mathbf{v} \times \mathbf{B} } +
\frac{1}{\mu_0\sigma} \del^2 \mathbf{B} \\
 \rho \der{ \mathbf{v} }{ t } & = & \mathbf{J} \times \mathbf{B}
  - \grad{p} + \rho \nu \del^2 \mathbf{v},
\end{eqnarray}
where $\mu_0$ is the vacuum magnetic permeability, $\sigma$ is the
fluid conductivity, $\rho$ is the fluid density, and $\nu$ is the
kinematic viscosity.  The time scale for magnetic diffusion is
\mbox{$\tau_\sigma = \mu_0 \sigma a^2$}, where $a$ is a characteristic
size of the system. Recast in dimensionless units, the induction
equation becomes
\begin{equation}\label{eq:induction}
\pder{ \mathbf{B} }{ t } = Rm \curl{ \mathbf{V} \times \mathbf{B} } +
\del^2 \mathbf{B},
\end{equation}
where the magnetic Reynolds number \mbox{$Rm = \mu_0 \sigma a v_0$} is
a measure of the rate of advection compared to the rate of magnetic
diffusion ($v_0$ is a characteristic speed).  The relative importance
of the Lorentz force in the flow dynamics is given by the interaction
parameter \mbox{$N=\sigma a B_0^2 / \rho v_0$}, where $B_0$ is a
characteristic field strength. For a kinematic dynamo, \mbox{$N \ll 1$}
so that the velocity field evolves independently of the magnetic
field. If the flow is stationary ({\it i.e.}, the flow geometry is
constant in time), the induction equation becomes linear in
$\mathbf{B}$; it can be solved as an eigenvalue equation by expanding
$\mathbf{B}$ as
\begin{equation}
\mathbf{B}(\mathbf{r},t) = \sum_i \mathbf{B}_i(\mathbf{r})
e^{\lambda_i t},
\end{equation}
where $\lambda_i$ are the growth rates of the magnetic eigenmodes
\mbox{$\mathbf{B}_i(\mathbf{r})$}. A dynamo is produced when at least
one eigenvalue has a positive real growth rate.

There are two fundamental requirements for a flow to produce a
dynamo. The first requirement is a sufficiently fast flow speed so
that the advection of the magnetic field overcomes ohmic
dissipation.\cite{Moffatt} There is a minimum magnetic Reynolds number
$Rm_{\rm crit}$ below which resistive diffusion dominates the
evolution of the field. As $Rm \rightarrow Rm_{\rm crit}$, it is
expected that the flow will more effectively amplify the initial seed
field to produce a dynamo. The second requirement is feedback---the
induced field must reinforce the initial seed field. Although the flow
may induce a strong response from an initial seed field, there is no
feedback to continue the cycle of magnetic field generation if the
induced field is perpendicular to the seed field.

The amplification and feedback can be quantified by the gain, defined
as
\begin{equation}\label{eq:gain}
{\rm gain} = \frac{ B_{\rm induced} \cos \delta + B_{\rm applied} }{
  B_{\rm applied} }.
\end{equation}
Here, $B_{\rm induced}$ is a measure of the amplitude of the induced
field, $B_{\rm applied}$ is a measure of the amplitude of the
applied field, and $\delta$ is the angle defining the relative
orientation of the induced field to the applied field. The mechanism
which produces a dynamo is described in terms of amplification and
feedback in Sec.\,\ref{sec:flow_modeling} and the use of
Eq.\,\ref{eq:gain} in analyzing measurements from the experiment is
described in Sec.\,\ref{sec:gain}.

\subsection{Experimental flow modeling}
\label{sec:flow_modeling}

\begin{figure*}
\includegraphics{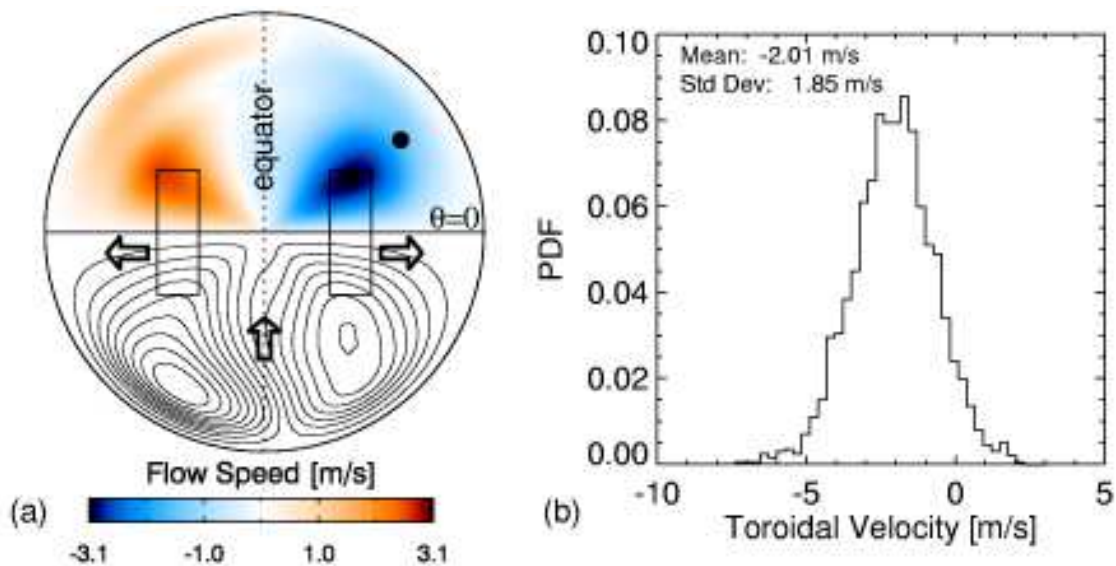}
\caption{(color)(a) Velocity field profiles fitted to LDV measurements
  of water flows generated by impellers identical to those used in the
  sodium experiment. The impeller rotation rate is 16.7\,Hz,
  corresponding to $Rm_{\rm tip}=100$ based on impeller tip speed. A
  contour plot of the toroidal flow is shown in the upper hemisphere,
  and poloidal stream lines are shown in the lower hemisphere. The
  poloidal flow is directed inward at the equator and outward at the
  poles. The two rectangles represent the location of the
  impellers. The $s_{1,0}$, $s_{2,0}$, $s_{4,0}$, $t_{1,0}$,
  $t_{2,0}$, and $t_{4,0}$ profiles are used for the flow model. (b)
  The probability distribution function for an LDV measurement of the
  toroidal velocity in the water model of the sodium experiment. The
  measurement location is shown as the circular dot in (a).}
\label{fig:flow}
\end{figure*}

Although the impeller-generated flow in the experiment is turbulent,
the mean flow can be approximated by a laminar model. Flows in a
spherical geometry are most easily modeled by the Bullard and Gellman
formalism.\cite{Bullard_and_Gellman_PTRSLA_1954} The velocity field is
described by a spherical harmonic expansion of toroidal and poloidal
stream functions\footnote{In a spherical geometry, toroidal field
lines are confined to spherical surfaces whereas poloidal field lines
penetrate spherical surfaces.}
\begin{multline}\label{eq:stream}
\mathbf{V}(r,\theta,\phi) = \sum_{\ell,m} \left( \del \times \del
  \times \left[ \frac{}{} s_{\ell,m}(r) Y_\ell^m(\theta,\phi)\
  \unit{r} \right] \right.\\ + \left. \del \times \left[ \frac{}{}
  t_{\ell,m}(r) Y_\ell^m(\theta,\phi)\ \unit{r} \right] \right)
\end{multline}
and the magnetic field is described by an expansion of flux functions
\begin{multline}\label{eq:B_expansion}
\mathbf{B}(r,\theta,\phi) = \sum_{\ell,m} \left(\del \times \del
  \times \left[ \frac{}{} S_{\ell,m}(r) Y_\ell^m(\theta,\phi)\
  \unit{r} \right] \right.\\ + \left. \del \times \left[ \frac{}{}
  T_{\ell,m}(r) Y_\ell^m(\theta,\phi)\ \unit{r} \right] \right),
\end{multline}
where the $Y_\ell^m(\theta,\phi)$ are the spherical harmonics,
$s_{\ell,m}(r)$ and $t_{\ell,m}(r)$ are radial scalar profiles
describing the poloidal and toroidal velocity field, and
$S_{\ell,m}(r)$ and $T_{\ell,m}(r)$ are radial scalar profiles
describing the poloidal and toroidal magnetic field. The
experimental flow is axisymmetric and is composed of primarily the
$t_{2,0}$ and $s_{2,0}$ terms. Hence, it is called a $t2s2$ flow.

The velocity field depicted in Fig.\,\ref{fig:flow}(a) is determined
from Laser Doppler Velocimetry (LDV) measurements performed in an
identical-scale water model of the sodium
experiment.\cite{Forest_et_al_MHD_2002} Liquid sodium at $120^\circ$C
has the same kinematic viscosity and density as water at $40^\circ$C,
so the flow measured in the water model should correspond to the
sodium flow. The magnitude of velocity fluctuations is extremely
large, as can be seen from the probability distribution function from
LDV measurements depicted in Fig.\,\ref{fig:flow}(b). The flow profile
depicted in Fig.\,\ref{fig:flow}(a) is therefore only realized in a
mean sense. The instantaneous flow profile can deviate significantly
from the mean, implying that the eigenmode growth rate should
fluctuate on the timescale of the flow evolution. The induced field
would then exhibit transient behavior, especially for flows near
$Rm_{\rm crit}$. The auto-correlation time of the LDV measurements of
the flow is \mbox{$\tau_c = 60 \pm 20$}\,ms for $Rm_{\rm tip} = 100$,
which serves as an estimate of the timescale for changes in the flow
profile.

\begin{figure}
\includegraphics{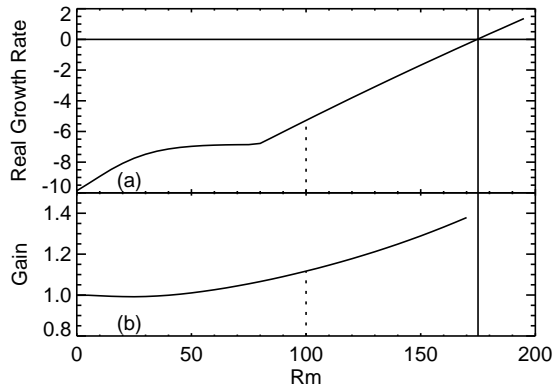}
\caption{(a) The kinematic growth rate of the dominant eigenmode
  versus $Rm$. The growth rate is calculated from the linearized
  induction equation using the flow profile constructed from LDV
  measurements in a water model of the sodium experiment, where
  $Rm_{\rm tip}=100$ ($Rm=98$ based on maximum velocity). It is
  assumed that the impellers generate the same flow profile as the
  impeller tip speed is increased and that the maximum flow speed
  scales linearly with the tip speed. (b) The corresponding gain as
  defined in Eq.\,\ref{eq:gain}.  The amplification of magnetic flux
  grows as the flow reaches $Rm_{\rm crit}=175$. Above $Rm_{\rm
  crit}$, the gain is undefined since the kinematic model does not
  account for the saturation of the magnetic field.}
\label{fig:growth_rate}
\end{figure}

\begin{figure*}
\includegraphics{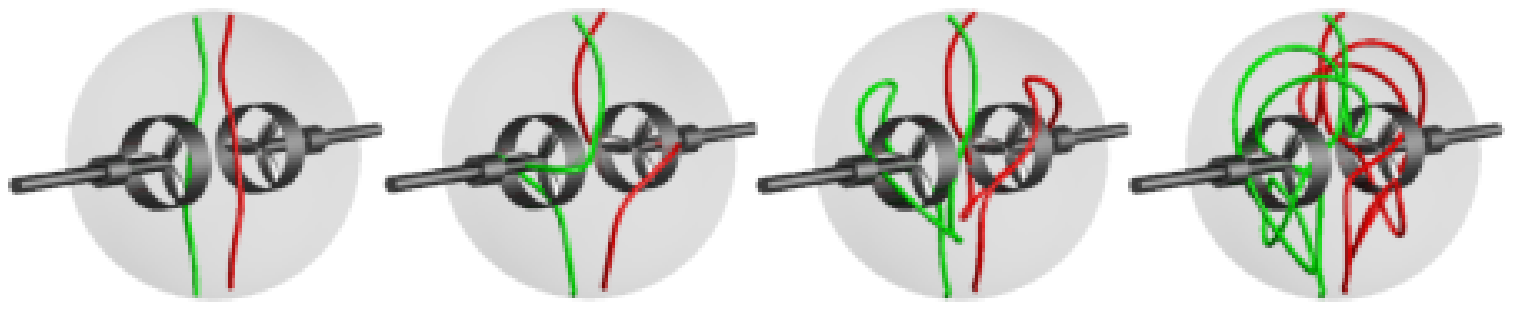}
\caption{(color) The mechanism of field generation in a laminar double-vortex
flow is modeled in the frozen flux limit ($Rm \rightarrow \infty$). A
field line directed through the equator is stretched by the poloidal
flow towards the pole and then twisted back onto itself by the
toroidal flow.}
\label{fig:stretch-twist-fold}
\end{figure*}

An eigenmode analysis of the flow profile shown in
Fig.\,\ref{fig:flow}(a) suggests that a magnetic field should be
generated for $Rm \ge 175$ as seen in
Fig.\,\ref{fig:growth_rate}(a). The structure of the magnetic
eigenmode with the largest growth rate for the $t2s2$ flow is
dominated by the $S_{1,1}$ term in the expansion of
Eq.\,\ref{eq:B_expansion}, corresponding to a dipole field oriented
perpendicular to the axis of symmetry of the flow (note the field
lines in the first panel of Fig.\,\ref{fig:stretch-twist-fold}). The
gain, shown in Fig.\,\ref{fig:growth_rate}(b), is calculated from
Eq.\,\ref{eq:gain}, where $B_{\rm induced}$ is the strength of the
transverse dipole field induced by the flow subjected to a transverse
dipole field of strength $B_{\rm applied}$. The azimuthal angle
between the induced and applied fields is $\delta$. It can be seen
from Fig.\,\ref{fig:growth_rate}(b) that the gain increases as $Rm
\rightarrow Rm_{\rm crit}$, though the feedback is insufficient to
produce a dynamo until $Rm > Rm_{\rm crit}$. It should also be noted
that Cowling's anti-dynamo theorem does not apply since the eigenmode
breaks the system's axisymmetry.\cite{Moffatt}

The physical mechanism of the gain and feedback for the $t2s2$ flow is
demonstrated by following the evolution of a field line in the
frozen-flux approximation (the diffusion term in
Eq.\,\ref{eq:induction} is neglected). In
Fig.\,\ref{fig:stretch-twist-fold}, two field lines parallel to the
equatorial plane are stretched to the poles (located where the drive
shafts enter the sphere) by the poloidal flow and twisted by the
toroidal flow back into their original direction.  The induced field
enhances the initial seed field, thereby increasing its magnetic
flux. This process continues as tension in the field lines builds. The
frozen-flux assumption eventually becomes invalid; the field line
tension presumably is relieved through magnetic reconnection due to
resistive diffusion. Once the magnetic field becomes sufficiently
strong, the interaction parameter becomes large. The Lorentz force
creates a torque on the flow, referred to as the back reaction, which
halts the field growth.

The laminar analysis above may be inadequate for describing the
experiment since turbulence also induces magnetic fields. Eddies in
the flow distort the magnetic field, creating small-scale magnetic
field structures which can affect the large-scale induced field. Mean
Field Theory suggests that velocity field fluctuations can enhance the
resistivity of the fluid by effectively increasing the rate of
diffusion (the $\beta$-effect),\cite{Reighard_and_Brown_2001} generate
large-scale currents through helical fluid motion (the
$\alpha$-effect),\cite{Parker_AJ_1955} and reduce the magnetic field
within the flow due to inhomogeneities in the magnitude of turbulent
fluctuations (the $\gamma$-effect).\cite{Krause_and_Raedler} The usual
assumptions of homogeneous, isotropic turbulence and scale separation
of fluctuations from the mean flow are not necessarily satisfied in
the experiment; hence the effects described are only used for
conceptual understanding.

\section{Description of the Experiment}
\label{sec:experiment}

The experiments are performed in a 1\,m diameter, 1.6\,cm thick,
stainless steel sphere shown in Fig.\,\ref{fig:experiment_schematic}.
Liquid sodium is transferred pneumatically from a storage vessel,
located in a vault beneath the floor, to the sphere using pressurized
argon. The sodium fills the sphere from the bottom until the liquid
level rises to an expansion tank connected to the top of the
sphere. The expansion tank accommodates changes in the volume of the
sodium due to variation in temperature. The weight of the sodium in
the storage vessel is monitored during the transfer to determine the
liquid level in the sphere. Electrical contact switches in the
expansion tank provide a redundant means of determining the sodium
fill level.

The conductivity of liquid sodium varies with temperature. It has a
maximum value of $1.5 \times 10^7(\Omega {\rm m})^{-1}$ near the
freezing point of sodium ($98^\circ$C) and decreases by about 4\% for
every $10^\circ$C\@. Since $Rm \propto \sigma$, the sphere is kept at
a temperature of 105--110$^\circ$C to optimize $Rm$ without freezing
the sodium. The sphere's temperature is maintained by a heat exchange
system which runs heat-transfer oil through a series of copper tubes
on the sphere's surface. The system provides 12\,kW of heating and
75\,kW of cooling to the oil. Kaowool insulation reduces the sphere's
ambient heat loss. The heat introduced to the sodium by the rotating
impellers is removed through the surface of the sphere by the heat
exchange system.

\begin{figure}
\includegraphics{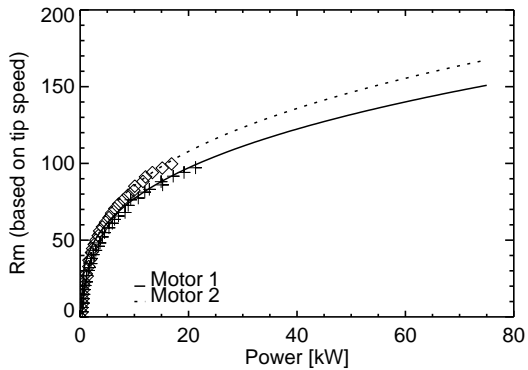}
\caption{The motor power follows the characteristic cubic relationship
with the impeller rotation rate which is proportional to $Rm$. The
curve is extrapolated to the maximum motor power available, yielding
$Rm_{\rm max}=150.$}
\label{fig:motorRamp}
\end{figure}

Two 30.5\,cm diameter impellers generate the flow. They are driven by
75\,kW motors controlled by variable frequency drives (VFDs). The
motors rotate the impellers at rates from 3--30\,Hz ($Rm_{\rm tip}$ is
between 18--180). Since $Rm_{\rm crit}$ depends on flow geometry, the
impellers have been modified to produce the desired flow: K{\"o}rt
nozzles limit the radial thrust and fins on the exterior of the
nozzles increase the mean toroidal flow.\cite{Forest_et_al_MHD_2002}
The magnetic Reynolds number $Rm_{\rm tip}$ is approximated using the
rotation rate of the impellers, which is measured by digital
encoders. Although convenient, this estimate is generous in that the
speed of the impeller tip is much larger than that of the bulk
flow. The VFDs record torque and power information. The motor power
follows a cubic relationship with rotation rate as shown in
Fig.\,\ref{fig:motorRamp}. Once a dynamo is achieved, it is expected
that the back reaction due to the Lorentz force will increase the
power required to drive the impellers beyond the cubic extrapolation
shown.

At atmospheric pressure, the rotating impellers cause cavitation by
creating a rapid drop in the local pressure near the impeller blades,
forming bubbles in the fluid. When the bubbles collapse against the
blades, they emit ultrasonic noise which is monitored by a transducer
mounted to the sphere. Since these bubbles disrupt the flow through
the impellers, the cavitation must be suppressed. This suppression is
accomplished by pressurizing the sphere with argon gas. The required
pressurization is determined empirically by increasing the sphere
pressure until the ultrasonic noise is minimized. For example, to
operate at a rotation rate of 20\,Hz, the sphere must be pressurized to
550\,kPa (80\,psi).

Two sets of coils generate magnetic fields used to study the inductive
response of the flow.  One set is coaxial with the drive shafts, the
other orthogonal, as shown in Fig\,\ref{fig:experiment_schematic}. A
DC power supply provides the coils with 600\,A to generate fields up
to 100\,G. The coils can produce dipole, quadrupole, transverse
dipole, and transverse quadrupole field configurations. For flow
speeds of 10\,m/s, the interaction parameter is $N=10^{-3}$; hence the
magnetic field is advected passively by the flow.

The magnetic field is measured using Hall-effect probes (Analog
Devices AD22151 Linear Output Magnetic Field Sensors) on integrated
circuits with internal temperature compensation. The probes saturate
at $\pm 170$ G. The signals are sent through a low-pass filter to
reduce the noise level. Since the sphere shields the probes from
frequencies higher than the skin-effect frequency, the low-pass filter
is designed to have a roll-off frequency that matches the skin-effect
frequency. For the $d=1.6$\,cm thick stainless steel sphere ($\eta
\sim 7.2\times 10^{-7}\,\Omega$m), this frequency is $f_{\rm skin}
\sim ( \pi \mu_0 \sigma d^2 )^{-1} \sim 1$\,kHz. Probes are positioned
on a grid on the surface of the sphere capable of resolving spherical
harmonic modes up to a polar order of $\ell=7$ and an azimuthal order
of $m=5$.  Linear arrays of probes in stainless steel tubes are
inserted radially into the sphere and are oriented to measure either
axial or toroidal magnetic fields. Data from the magnetic probes is
sampled by 16-bit digitizers on PC-based data acquisition cards at a
rate of 1\,kHz per channel, which is sufficient to resolve
fluctuations due to eddies down to the resistive dissipation scale
(see Sec.\,\ref{sec:turbulence}). The stainless-steel tubes encasing
the internal sensor arrays vibrate when the impellers are driven at
rotation rates above 15\,Hz ($Rm_{\rm tip}=90$). Since the amplitude
of the vibrations increases with flow speed, experiments are limited
to the lower rotation rates to prevent damaging the tubes and risking
a breach. A comparison of data from experiments before the internal
arrays were installed with data from experiments with the tubes
indicates that the disturbance in the flow due to the tubes has
negligible effect on the large-scale induced magnetic field. Future
experiments will be performed without the internal sensor arrays to
reach higher rotation rates.

\begin{figure*}
\includegraphics{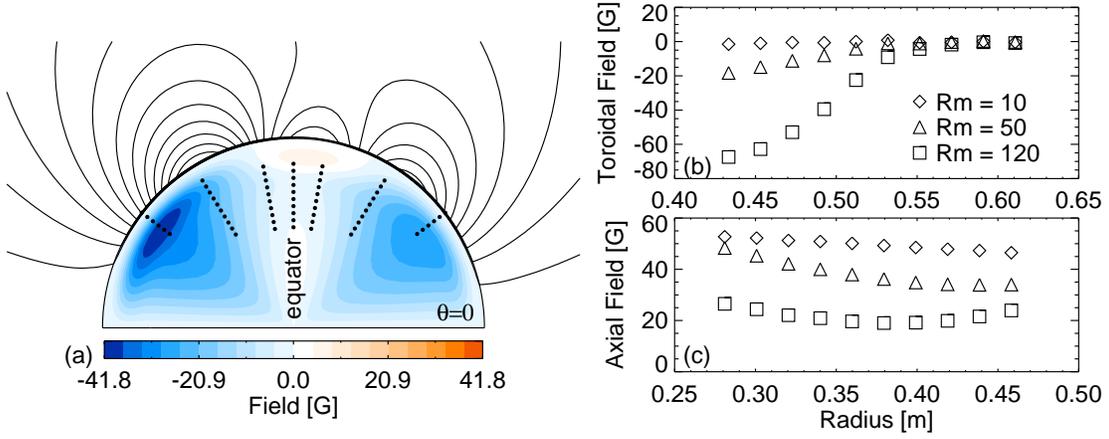}
\caption{(color) (a) A plot of the toroidal magnetic field produced by
  the $\omega$-effect when a 60\,G axial field is applied to a $t2s2$
  flow with $Rm=100$. The contours are calculated by fitting the
  expected $T_{\ell,m}(r)$ profiles to measurements of the internal
  toroidal field. The dots indicate the position of the Hall
  probes. External field lines are shown based on a vacuum-field
  expansion fit to measurements from the surface probe array. (b)
  Toroidal magnetic field measurements near the toroidal maximum above
  the impeller and (c) axial magnetic field measurements in the
  equatorial plane for various $Rm$.}
\label{fig:fieldcontour}
\end{figure*}

\section{Magnetic Field Measurements}
\label{sec:magneticfield}

\subsection{Reconstruction of the mean magnetic field}
\label{sec:meanfield}

The induction effects are studied by applying an axial magnetic field
to both flowing and stationary sodium and comparing the measured
fields. The flow's differential rotation wraps the field lines around
the drive shaft axis to produce a toroidal magnetic field through the
so-called $\omega$-effect, as seen in
Fig.\,\ref{fig:fieldcontour}(a). The contour plot is generated by
fitting the coefficients of the harmonic expansion in
Eq.\,\ref{eq:B_expansion} to toroidal field measurements (up to order
\mbox{$\ell=3$}). The $\omega$-effect is very efficient at amplifying
the applied magnetic field; measurements of the toroidal magnetic
field shown in Fig.\,\ref{fig:fieldcontour}(b) indicate that the
induced toroidal field increases with $Rm_{\rm tip}$ and is larger than the
applied field for \mbox{$Rm_{\rm tip}=120$}. In contrast,
Fig.\,\ref{fig:fieldcontour} shows that the axial magnetic field in
the equatorial plane is reduced by half. This reduction of poloidal
magnetic flux can be explained by the effect of flux expulsion due to
the strong poloidal circulation in the $t2s2$ flow.\cite{Moffatt}

The mean field structure shown in Fig.\,\ref{fig:fieldcontour}(a) is
obtained by an averaging procedure which calculates the peak of the
histogram of magnetic field measurements from each probe. This
technique is necessary since some of the signals have non-Gaussian
statistics (discussed in Sec.\,\ref{sec:turbulence}). The resulting
mean field is predominantly due to odd harmonics in accordance with
the selection rules governing the interaction terms of the induction
equation.\cite{Bullard_and_Gellman_PTRSLA_1954}

\begin{figure}
\includegraphics{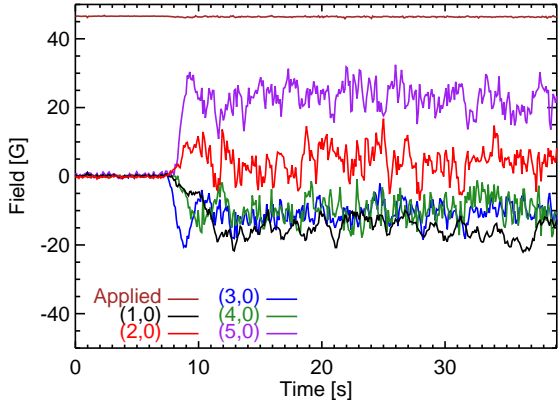}
\caption{(color) Time series plots of the axisymmetric modes evaluated
  at $\theta=0$. A constant external dipole field is applied and the
  motors are turned on at $t=7$s.}
\label{fig:mode_timeseries}
\end{figure}

The magnetic field outside the sphere is reconstructed from the field
harmonics calculated from the surface array measurements. The poloidal
flux lines are shown in Fig.\,\ref{fig:fieldcontour}(a). Since the
Hall probes on the surface of the sphere lie outside regions
containing currents, the measured field can be described in terms of
a vacuum field potential which satisfies $\nabla^2\Phi = 0$, where
$\mathbf{B} = - \grad{\Phi}$. Solving Laplace's equation for the
potential in a spherical geometry yields
\begin{equation}\label{eq:expansion}
\Phi(r,\theta,\phi) = \sum_{\ell,m} \left( C_{\ell,m} r^{\ell} +
\frac{ D_{\ell,m} }{ r^{\ell + 1} } \right) Y_\ell^m(\theta,\phi).
\end{equation}
The $C_{\ell,m}$ terms are due to currents in the external field coils
whereas the $D_{\ell,m}$ terms are due to currents in the flow. The
field can be separated into applied and induced fields whose radial
components are
\begin{eqnarray}\label{eq:applied}
\mathbf{B}_{\rm applied} \cdot \unit{r} & = & - \sum_{\ell,m}
C_{\ell,m} r^{\ell-1} \ell\, Y_\ell^m(\theta,\phi),\\
\label{eq:response}
\mathbf{B}_{\rm induced} \cdot \unit{r} & = & \sum_{\ell,m} \frac{ D_{\ell,m}
}{ r^{\ell + 2} } (\ell+1) Y_\ell^m(\theta,\phi).
\end{eqnarray}
The external measured magnetic field can be completely described in
terms of these expansion coefficients. To determine the coefficients
$D_{\ell,m}$ for a given set of magnetic field measurements, the
applied field is subtracted from the measured field, and a design
matrix $\mathsf{A}_{ij}$ is constructed which satisfies $\mathsf{B}_i
= \mathsf{A}_{ij} \mathsf{D}_j$, where $\mathsf{B}_i =
B_r(r_i,\theta_i,\phi_i)$ is an array of measurements of the induced
field and $\mathsf{D}_j = D_{\ell_j,m_j}$ is the array of expansion
coefficients in Eq.\,(\ref{eq:response}). An example of an element of
the design matrix is
\begin{equation}
A_{i,j} = \left( \frac{\ell_j+1}{r_i^{\ell_j+2}} \right)
Y_{\ell_j}^{m_j}(\theta_i,\phi_i).
\end{equation}
The coefficients are obtained by matrix inversion using Singular Value
Decomposition\cite{NumericalRecipies} to solve $\mathsf{D}_j =
(\mathsf{A}_{ij})^{-1} \mathsf{B}_i$. Axisymmetric mode time series
are shown in Fig.\,\ref{fig:mode_timeseries}, where the modes have been
evaluated at $\theta=0$. The dipole component of the induced magnetic
field grows non-linearly with $Rm_{\rm tip}$ and cannot be explained by
induction effects from an axisymmetric flow. The dipole response
results from a turbulent EMF generated by correlated turbulent
fluctuations in the flow.\cite{Spence_EMF_2006}

\subsection{Measuring gain with a transverse applied field}
\label{sec:gain}

\begin{figure}
\includegraphics{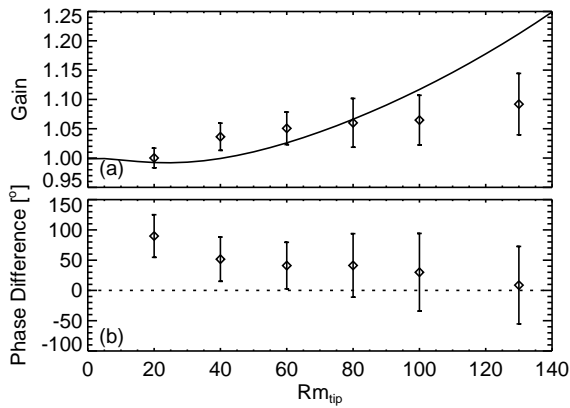}
\caption{(a) Amplification of a magnetic field applied perpendicular
  to the drive shaft axis as a function of $Rm_{\rm tip}$. Gain is defined in
  Eq.\,\ref{eq:gain}. The solid line is the gain predicted by the
  kinematic model shown in Fig.\,\ref{fig:growth_rate}(b). (b)
  Azimuthal angle between the induced transverse dipole field and the
  applied field. Error bars on both plots indicate RMS fluctuation
  levels.}
\label{fig:gain}
\end{figure}

The proximity of the flow to magnetic self-excitation is studied by
applying a magnetic field perpendicular to the drive shaft
axis. Recall from Sec.\,\ref{sec:kinematicdynamo} that the anticipated
structure of the magnetic field generated by the dynamo is a dipole
oriented in this direction. Figure~\ref{fig:growth_rate}(b) suggests
that such a field should be amplified by the flow. As in
Sec.\,\ref{sec:kinematicdynamo}, the gain in Eq.~\ref{eq:gain} is
calculated using the $\ell=1$, $m=1$ expansion term in
Eq.\,\ref{eq:response} for $B_{\rm induced}$, and $\delta$ is the
azimuthal angle between the applied and induced fields. The gain is
less than one if a component of the induced field is anti-parallel to
the applied field, indicating attenuation, and is greater than one if
the fields are in phase.  The gain is determined from measurements of
the induced magnetic field and is shown to increase with $Rm_{\rm
tip}$ in Fig.\,\ref{fig:gain}(a), though not as quickly as anticipated
from the kinematic model. Figure~\ref{fig:gain}(b) shows that the
induced field is somewhat out of phase with the applied field at low
$Rm_{\rm tip}$ and that the alignment improves as $Rm_{\rm tip}$
increases. These measurements will be used to obtain a flow profile in
the sodium experiment which maximizes the gain.

\section{Properties of MHD Turbulence}
\label{sec:turbulence}

\begin{figure*}
\includegraphics{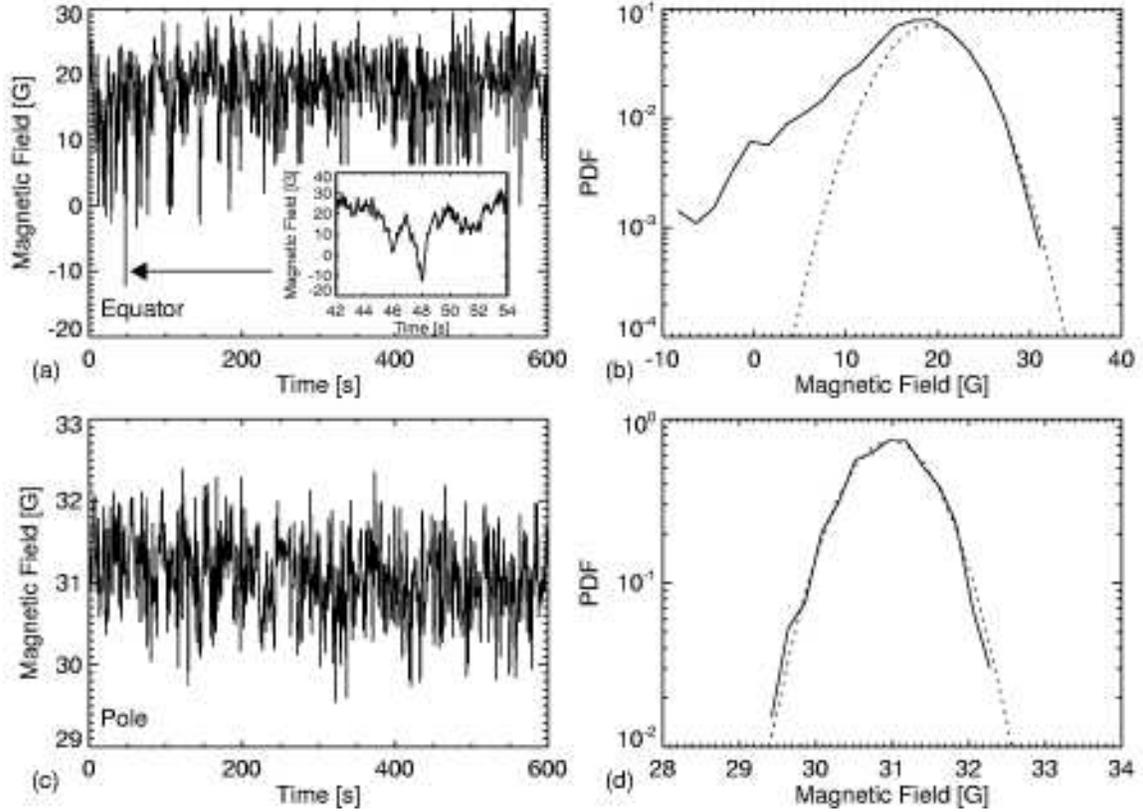}
\caption{(a) Measurement from a single Hall probe on the surface of
  the sphere near the equator. (b) The probability distribution for
  the signal in (a). The impeller rotation rate for this example is
  16.7\,Hz ($Rm=100$) and a 60\,G dipole field is applied. A Gaussian
  fit to the right side of the distribution is shown (dotted line) to
  illustrate the asymmetry. (c) The time series from a probe near the
  drive shaft axis (or pole) and (d) its probability distribution is
  shown for comparison.}
\label{fig:TimeSeries}
\end{figure*}

\begin{figure}
\includegraphics{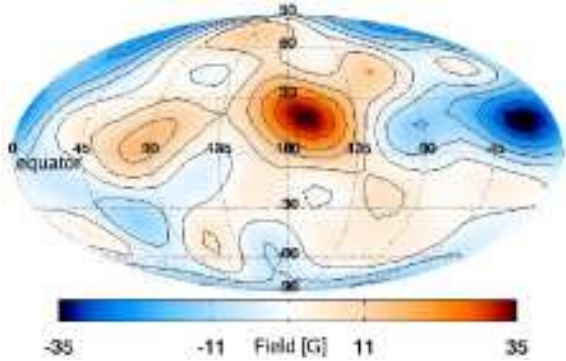}
\caption{(color) The mode structure of the intermittent bursts
  corresponds to a dipole field aligned perpendicular to the drive
  shaft axis, as seen from the surface magnetic field (the axis of
  rotation is vertical). The mean magnetic field has been removed to
  isolate the structure of the fluctuation.}
\label{fig:event_structure}
\end{figure}

\begin{figure}
\includegraphics{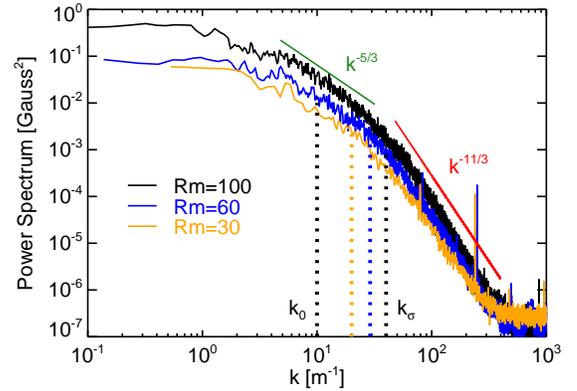}
\caption{(color) The inferred spatial spectrum constructed from the
 Fourier transform of the signal from a Hall probe inside the sphere
 for three different values of $Rm_{\rm tip}$. The probe is located
 just above the impeller near the toroidal maximum in
 Fig.\,\ref{fig:fieldcontour}. Fluctuations are assumed to be due to
 convection of spatial variations in the field. The dispersion
 relation is $\omega = k v_0$ where $v_0$ is determined from velocity
 measurements in the water model of the sodium experiment. The
 wavenumber corresponding to the scale size of the experiment is
 labeled as $k_0$. The wavenumber corresponding to the dissipation
 scale $k_\sigma$ is labeled for the three spectra.}
\label{fig:PowerLaw}
\end{figure}

The preceding analysis documents the large-scale induced mean
field. Since the flow is turbulent with $Rm \gg 1$ and the magnetic
field is relatively weak, eddies can twist and stretch field lines at
many scales. While smaller eddies are advected by the mean flow,
larger eddies can distort the large-scale flow itself and cause
intermittent behavior in the induced
field.\cite{Tennekes_and_Lumley} Figure~\ref{fig:TimeSeries} shows
signals from two Hall probes on the surface of the sphere and their
probability distribution functions (PDFs). In addition to random
fluctuations with a Gaussian distribution, the Hall probe signal in
Fig.\,\ref{fig:TimeSeries}(a) has large-amplitude bursts which skew
its PDF shown in Fig.\,\ref{fig:TimeSeries}(b). The time average of
the signal for this probe is therefore not the most-probable value of
the field. The signal shown in Fig\,\ref{fig:TimeSeries}(c) from a
probe near one of the drive shafts, however, has normal
statistics. Examination of the structure of the magnetic field during
the bursts reveals a strong $m=1$ mode shown in
Fig.\,\ref{fig:event_structure}. The turbulent flow thus induces an
axisymmetric field on average punctuated by intermittent bursts that
break the symmetry.

Eddies at the smaller scales create structure in the magnetic field
down to the viscous dissipation scale.\cite{Kraichnan_1967} This
structure is evident in the spatial spectrum in
Fig.\,\ref{fig:PowerLaw}.  The spatial spectrum is constructed from
the power spectrum of the signal from a Hall probe that measures the
toroidal field just above one of the impellers. The $k^{-5/3}$ scaling
can be derived from the induction equation using a weak-field
approximation. The magnetic field is described in terms of a mean
and a weak fluctuating part, $\mathbf{B} = \mathbf{B}_0 +
\widetilde{\mathbf{B}}$. The Fourier transform of
Eq.\,\ref{eq:induction} becomes
\begin{equation}
\label{eq:FourierInduction}
\left( i\omega + \frac{k^2}{\mu_0\sigma} \right)
\widetilde{\mathbf{B}}_{{\mathbf{k},\omega}} = i \mathbf{k} \times
\mathbf{v}_{\mathbf{k},\omega} \times \mathbf{B}_0,
\end{equation}
where second-order advection terms have been assumed to be negligible
due to the weak-field approximation. Magnetic fluctuations are due
primarily to the advection of eddies by the mean flow (the Taylor
hypothesis\cite{taylor:476}) and so the dispersion relation is
approximately $\omega(k) \sim k v_0$. Dividing
Eq.\,\ref{eq:FourierInduction} by this dispersion relation gives
\begin{equation}
\label{eq:scaling}
\left( \frac{i\omega}{k v_0} + \frac{k}{\mu_0\sigma v_0} \right)
\widetilde{\mathbf{B}}_{{\mathbf{k},\omega}} = i \frac{\mathbf{k}}{k} \times
\frac{\mathbf{v}_{\mathbf{k},\omega}}{v_0} \times \mathbf{B}_0.
\end{equation}
At scales $k \ll k_\sigma \equiv \mu_0 \sigma v_0 = Rm / a$, the
dissipation term is small. Thus, magnetic fluctuations at these scales
are primarily due to advection of the mean field by velocity
fluctuations, giving $\widetilde{B}_{k,\omega}^2 / B_0^2 \sim
v_{k,\omega}^2 / v_0^2$.  For isotropic, homogeneous turbulence, the
velocity spectrum is the Kolmogorov spectrum $E_K(k) = v_{k,\omega}^2
/ k \propto k^{-5/3}$. The resulting magnetic spectrum is $E_M(k) =
\widetilde{B}_{k,\omega}^2 / k \propto k^{-5/3}$ as seen in
Fig.\,\ref{fig:PowerLaw} for the range $k_0 < k < k_\sigma$. This
scaling was observed in the Maryland dynamo
experiment,\cite{Peffley_and_Cawthorne_and_Lathrop_PRE_2000} but in
the von K\'arm\'an sodium (VKS) experiment, a $k^{-1}$ scaling was
observed.\cite{bourgoin:3046} The discrepancy was attributed to a saturation of the induction
mechanism.

For the range $k \gg k_\sigma$, the advection and diffusion terms
become comparable. From Eq.\,\ref{eq:scaling}, we have the scaling
$\widetilde{B}_{k,\omega}^2 / B_0^2 \sim \left( \mu_0 \sigma / k
\right)^2 v_{k,\omega}^2$. Hence, the magnetic spectrum is $E_M(k)
\propto k^{-2} E_K(k) \propto k^{-11/3}$ in the resistive dissipation
range, as seen in Fig.\,\ref{fig:PowerLaw} for $k > k_\sigma$. The
$k^{-11/3}$ scaling was observed in the VKS experiment, but the
Maryland experiment observed a steeper spectrum due to shielding
effects of the stainless-steel vessel.

\begin{table}[h]
\caption{The resistive dissipation scale determined from the spatial
spectrum in Fig.\,\ref{fig:PowerLaw}. The velocity is measured in the
water model of the experiment at the probe's position.}
\begin{ruledtabular}
\begin{tabular}{ddd}
\multicolumn{1}{c}{$Rm_{\rm tip}$} & 
\multicolumn{1}{c}{$v_0\,[\mbox{m/s}]$} &
\multicolumn{1}{c}{$k_\sigma\,[\mbox{m}^{-1}] $}\\
\hline
30 & 0.39 & 20\\
60 & 0.78 & 29\\
100 & 1.2 & 40\\
\end{tabular}
\end{ruledtabular}
\label{tab:diss_scale}
\end{table}

The dissipation scale $k_\sigma$ is evident from the knee in the wave
number spectrum of Fig.\,\ref{fig:PowerLaw}. The results are
summarized in Tab.\,\ref{tab:diss_scale}, showing that $k_\sigma$
increases with $Rm_{\rm tip}$. Consequently, the magnetic field gains
structure at smaller scales as $Rm_{\rm tip}$ increases, down to scale
sizes of \mbox{$\ell_\sigma = 2\pi/k_\sigma = 16$\,cm} at $Rm_{\rm
tip}=100$. A magnetic Reynolds number for the turbulent part of the
flow can be constructed assuming that \mbox{$k_\sigma = Rm_{\rm turb}
k_0$}. Using the values in Tab.\,\ref{tab:diss_scale}, it can be shown
that $Rm_{\rm turb} = 0.2 Rm_{\rm tip}$. This relation implies that
the small-scale turbulent eddies produced by the mean flow are slower
than the injection-scale eddies. The dissipation scale is therefore
only about one decade smaller than the injection scale rather than the
two decades expected for a flow with \mbox{$Rm \sim 100$}. As a
result, the small-scale dynamics of the flow has little role in
generating magnetic fields; it is the largest eddies in the flow which
induce magnetic fields and which have the greatest impact on the
transition to a dynamo.

\section{Conclusion}
\label{sec:conclusion}

The magnetic field induced by a turbulent flow of liquid sodium
demonstrates flux expulsion and the $\omega$-effect. The flow
generated in the experiment induces a field that provides
amplification and feedback, the necessary ingredients of a
dynamo. Power spectrum measurements show that the resistive
dissipation scale is inversely proportional to $Rm$. The turbulent
flow creates intermittent bursts of large-scale magnetic fields that
have a structure similar to the magnetic field predicted from the
laminar kinematic dynamo model. Consequently, as a flow nears $Rm_{\rm
crit}$, the transition to a dynamo appears to be intermittent rather
than smooth.

Special thanks go to C.~A. Parada, B.~A. Grierson, and M.~Fix for
their aid in constructing the experiment and to R.~A. Bayliss for
helpful conversations. The authors would like to acknowledge the
generous financial support for this project from the Department of
Energy, the National Science Foundation, the Packard and Sloan
Foundations, and the Research Corporation.

\end{document}